\documentclass[preprint,prb,aps,showpacs]{revtex4}

\usepackage[T1]{fontenc}
\usepackage{setspace}
\usepackage{graphics}
\usepackage{amssymb}
\usepackage{color}

\begin{document}

\def\beq{\begin{equation}}
\def\eeq{\end{equation}}


\title{Local order and orientational correlations in liquid and crystalline phases of carbon tetrabromide from neutron powder diffraction measurements}

\author{L. Temleitner\footnote{Present address: Japan Synchrotron Radiation Research Institute (JASRI, SPring-8), 1-1-1 Kouto, Sayo-cho, Sayo-gun, Hyogo 679-5198, Japan}}
\email{temla@szfki.hu}
\author{L. Pusztai}
\affiliation{Research Institute for Solid State Physics and Optics,\\
P.O.Box 49, H-1525 Budapest, Hungary}



\begin{abstract}

The liquid, plastic crystalline and ordered crystalline phases of CBr$_4$ were studied using neutron powder diffraction. The measured total scattering differential cross-sections were modelled by Reverse Monte Carlo simulation techniques (RMC++ and RMCPOW). Following successful simulations, the single crystal diffraction pattern of the plastic phase, as well as partial radial distribution functions and orientational correlations for all the three phases have been calculated from the atomic coordinates ('particle configurations'). The single crystal pattern, calculated from a configuration that had been obtained from modelling the powder pattern, shows identical behavior to the recent single crystal data of Folmer et al. (Phys. Rev. {\bf B77}, 144205 (2008)). The BrBr partial radial distribution functions of the liquid and plastic crystalline phases are almost the same, while CC correlations clearly display long range ordering in the latter phase. Orientational correlations also suggest strong similarities between liquid and plastic crystalline phases, whereas the monoclinic phase behaves very differently. Orientations of the molecules are distinct in the ordered phase, whereas in the plastic crystal their distribution seems to be isotropic.

\end{abstract}


\maketitle

\section{Introduction}
\label{intro}

Carbon tetrabromide (CBr$_4$) is a model material of crystalline solids of tetrahedral molecules that, on raising the temperature, shows an ordered-disordered crystal phase transition. At ambient pressure it has two solid modifications, as well as its liquid and gaseous phases. The phase transitions occur at 320 K (ordered crystal-plastic crystal), 365 K (solid-liquid) and 462 K (liquid-gas)\cite{More1977II,CRCHandbook}. The low temperature ordered (II, $\beta$-) phase consists of monoclinic ($C2/c$) cells, whose asymmetric unit contains 4 molecules\cite{More1977II}. The higher temperature plastic (orientationally disordered, I, $\alpha$-) phase is face centered cubic ($Fm\bar{3}m$), where only the centers of the molecules maintain the translational symmetry\cite{More1977I}. At higher pressures other solid phases exist and one of them appears to be plastic\cite{Anderson}, which is recently studied by neutron diffraction\cite{Levit}; in what follows, the terms 'ordered crystalline' and 'plastic (disordered) crystalline' will refer only to the ambient pressure modifications.

The scientific interest has been mainly concentrated on the plastic crystalline phase, where the molecules possess higher symmetry than their lattice site symmetry\cite{DollingMP}. Fulfilling the crystallographic site symmetry in time average, the molecules become rotating\cite{More1977I}.
Due to this phenomenon some macroscopic properties become similar to those found in the liquid state, e.g. the thermal resistivity is almost as large as in the liquid state, showing that the mean free path of elastic waves become short\cite{Anderson}. At the microscopic level, structural and dynamic properties have been studied extensively using (powder\cite{DollingMP, DollingMCLC}, single crystal\cite{More1977I, MoreJPC1980a,Folmer}) diffraction and triple-axis spectrometry\cite{MoreJPC1980a, MoreDFC, More1984}. Numerous models\cite{More1977I, DollingMCLC, DollingMP, MoreJPC1980a, MoreJPC1980b, MoreDFC, More1984, Hohlwein, Dove1, Dove2, Folmer} have been invented for describing the scattering pattern from this phase, taking into account more and more detailed effects as computer power has been increasing. An important effect in the static (or snapshot) picture is the steric hindrance due to repulsion between bromine atoms of neighboring molecules\cite{MoreJPC1980b}. Simulations fulfilling this condition\cite{MoreJPC1980b, Dove1, Folmer} have provided different results for the orientational probability in relation to the unit cell; simulations based on a Frenkel model with 6 orientations\cite{MoreJPC1980b, Folmer} provided more ordered real space structures than molecular dynamics calculations\cite{Dove1}.

The ordered phase has been studied in relation to the order-disorder transition and in comparison with similar materials, using diffraction methods\cite{More1977II, Powers, Negrier, Folmer}, measurements of thermodynamic parameters\cite{Michalski} and via molecular dynamics simulations\cite{Zielinski}. The structure is a distorted face centered cubic one, which could eventually be refined as monoclinic\cite{More1977II}. 

Liquid diffraction data was first published in 1979\cite{DollingMP}; the first discussion appeared only in 1997\cite{Bako}, based on Reverse Monte Carlo\cite{Pusztai88} (RMC) structural modelling. The authors   found that, in their system of rigid molecules, both molecular center -- molecular center and bromine atom -- bromine atom correlations resemble to those present in a closely packed structure. Interestingly it was also suggested that 'the packing density is such that the molecules have interlocking structures and cannot rotate freely'; this statement seems to oppose common sense expectations.

The similarity between the total scattering structure factor of the liquid and the diffuse scattering part of the total powder diffraction patterns of the plastic phase could be spotted for some halomethanes\cite{Pardo2007}. Molecular dynamics simulation of Rey\cite{Rey2009liq} and RMC modelling of our research group\cite{SzilviXCl4} presented similar orientational correlations on the class of tetrahedral shape molecular liquids. Rey recently published\cite{Rey2008plastic} a comparison study between plastic, liquid, gaseous phase of CCl$_4$ and neopentane, which suggests the short range orientational order remains from liquid to plastic crystal phase transition, but long range orientational correlations appear due to translation symmetry. For CBr$_4$ some authors\cite{DollingMP, Dove1, Bako} have also suggested an analogy between the liquid and plastic crystalline phases, but this comparison has not yet been made in detail.

The present work focuses on changes of the extent of order/disorder in different phases of carbon tetrabromide, by means of neutron (powder) diffraction and subsequent Reverse Monte Carlo modelling.
In section \ref{expt} we present the measured total powder diffraction patterns of CBr$_4$ in the two solid and in the liquid phases. Section \ref{rmc} describes variants of RMC modelling as applied to liquid and crystalline systems, together with details of calculations carried out during the present investigation. In Section \ref{results} results of analyses of particle configurations provided by RMC modelling are presented and discussed, whereas Section \ref{conclusions} summarizes our main findings.

\section{Experiment}
\label{expt}

\begin{figure}[p]
\begin{center}
\rotatebox{0}{\resizebox{0.4\textwidth}{!}{\includegraphics{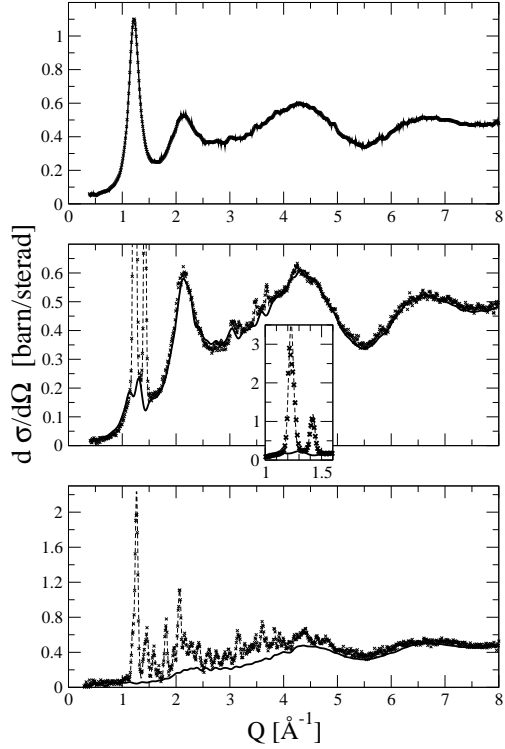}}}
\caption{\label{fig_expsq}
Measured and simulated powder diffraction patterns of CBr$_4$ at 298~K (lower panel, ordered crystalline phase), 340~K (middle panels, plastic crystalline phase) and 390~K (upper panel, liquid phase). Crosses: measured differential cross-section; solid line: RMC calculated diffuse intensities; dashed line: RMC calculated total scattering intensities.
}
\end{center}
\end{figure}

Neutron diffraction measurements have been carried out using the SLAD diffractometer\cite{slad} at the former Studsvik NFL in Sweden. At a wavelength of 1.119~\AA, the experiment was carried out at temperatures 298~K, 340~K and 390~K and at ambient pressure over the momentum transfer range of 0.29 -- 10.55~\AA$^{-1}$. The powdered sample was sealed in an 8~mm thin-walled vanadium can and standard furnace was used for measurements above room temperature. In the 'total scattering' type experiment, scattered intensities from the sample, empty can (+furnace), instrumental background  and standard vanadium rod were recorded. A standard normalization and correction (for absorption, multiple and inelastic scattering) procedure\cite{howecorr} has been applied using the \texttt{CORRECT} program\cite{correct}. Corrected and normalized datasets\cite{measured} are shown in FIG.\ref{fig_expsq}.

\section{Reverse Monte Carlo modelling}
\label{rmc}

\subsection{Reverse Monte Carlo modelling of crystalline powder samples}

The Reverse Monte Carlo simulation procedure\cite{Pusztai88} is a useful tool for gaining a deeper 
understanding and a better interpretation of diffraction data than it could be achieved by using direct methods. The RMC algorithm provides sets of 3 dimensional particle coordinates ('configurations') which are consistent with experimental (mainly diffraction) results. During the procedure, coordinates of the particles in the configuration are changed so that the measured datasets are approached by the simulated ones within experimental errors. For a detailed description, see Refs.\cite{Pusztai88, McGreevy2001, rmc++}. 

The computation path from the particle coordinates to the simulated diffraction dataset differs for the cases of liquid (or amorphous) and crystalline states. Liquids and amorphous materials can be considered isotropic beyond nearest neighbor distances so that in real and in reciprocal space, a one dimensional formalism is widely used. From the particle coordinates, partial radial distribution functions ($g_{xy}(r)$, prdf) can be calculated easily. They can be Fourier-transformed and weighted for the actual experiment, thus providing the total scattering structure factor ($F(Q)$) which is an experimental quantity:
\begin{eqnarray}
F(Q)&=&\frac{d\sigma}{d\Omega} - \sum_x c_x\frac{\sigma_x}{4\pi} = \sum_{x,y} c_x c_y f_x(Q) f^*_y(Q) \times \nonumber \\
&\times& \rho \int_0^{\infty} 4\pi r^2 \left( g_{xy}(r)-1\right) \frac{\sin Qr}{Qr} dr,
\end{eqnarray}
where $\frac{d\sigma}{d\Omega}$, $\sigma_x$, $c_x$, $f_x(Q)$ and $\rho$ denote the differential cross-section, the scattering cross-section, concentration, form factor (or scattering length) of the atom type $x$ and the atomic number density of the sample, respectively. This method is implemented by the \texttt{RMC++}\cite{rmc++} (and previously, by the \texttt{RMCA}\cite{rmca}) software package.

In the case of crystals, where the long (Bragg-peaks) and short range order (diffuse scattering) appear simultaneously, different approaches exist, depending on the available experimental $Q$-range. Wide $Q$-range is needed in the cases where the radial distribution function is used as experimental data to be fitted, in order to reduce Fourier-errors in the rdf. Fitting to the rdf occurs in the \texttt{PDFfit}\cite{pdffit} and the \texttt{RMCProfile}\cite{rmcprofile} methods. The former is used to fit only to the rdf, whereas the latter applies $Q$-space refinement to the convoluted structure factor for the Bragg-peaks, as well (convolution of the experimental data with a step function corresponding to the simulation box size is necessary to avoid the finite configuration cell effect). Although \texttt{PDFfit} provides results faster than \texttt{RMCProfile}, the latter is able to capture more detailed structural information obtained from modelling also in $Q$-space.

In contrast, the \texttt{RMCPOW}\cite{rmcpow} method fits the measured differential cross-section, both the Bragg- and diffuse-scattering parts, in $Q$-space. It uses the supercell approximation where the configuration cell is the repetition of the unit cell in each direction. For obtaining the structure factor in the reciprocal space a 3 dimensional Fourier-transformation is needed using the coordinates ($\vec{R}_j$) of each atom:
\begin{equation}
F(\vec{q})=\sum_{j=1}^{N} f_j(q) \exp \left( i \vec{q}\vec{R}_j\right).
\end{equation}
The coherent part of the powder diffraction cross-section ($\frac{d \sigma_c}{d \Omega}$) can be calculated from the structure factors as
\begin{equation}\label{eq:cs}
\frac{d \sigma_c}{d \Omega} =\frac{2 \pi^2}{NV} \sum_{\vec{q}} \frac{F(\vec{q}) F^*(\vec{q}) \delta (Q-q)}{q^2},
\end{equation}
where $N$, $V$, $\vec{q}$, $Q$ denote the number of atoms in the unit cell, the volume of the unit cell, an allowed (by the configuration supercell) reciprocal lattice vector and the modulus of the observable scattering vector, respectively. \texttt{RMCPOW} handles supercell intensities as Bragg-reflections if a given point is the reciprocal lattice point of the unit cell; otherwise the intensity at that given point contributes to the diffuse scattering intensity. Diffuse intensities (which are assumed to vary smoothly) are locally averaged in the reciprocal cell and finally summed up into a $|\vec{Q}|$ histogram. For Bragg-intensities the same summation is performed (without averaging), and after that the instrumental resolution function (instead of $\delta$ distribution in EQ. \ref{eq:cs}) is applied to them.

Although \texttt{RMCPOW} needs the largest computational effort of the three methods to make a Monte Carlo move, there are some advantages.  First, there is no need to convolute the original dataset with anything related to the calculation itself. Furthermore, a too wide $Q$-range is not necessary for three reasons: (i) in crystallography, the low $Q$-range is exploited for determining the average structure, due to the fact that thermal displacements decrease the Bragg-intensities with increasing $Q$. (ii) The molecular structure is often known, at least approximately, so it can simply be built in the calculation via constraints. (iii) Short range order (intermolecular) correlations have a significant contribution in reciprocal space\cite{orsi} below 6-10~\AA$^{-1}$. 
(Note also that a wide $Q$-range necessitates much more computational time.) Hence, \texttt{RMCPOW} makes the examination of the local order possible from (total scattering type) powder diffraction measurement(s) on laboratory x-ray\cite{biologusok} machines and on neutron diffractometers at medium power reactor sources.

In the \texttt{RMCPOW} and \texttt{RMC++} programs, real space constraints, including coordination number constraints, are also available. Since (crystalline and liquid) CBr$_4$ may contain intermolecular BrBr correlations in the intramolecular region, coordination number constraints are not the best tools for keeping molecules together. To avoid this problem, during the present research the fixed neighbor constraint\cite{rmc++} (FNC) concept, which had been available already in \texttt{RMC++}, has been implemented in the \texttt{RMCPOW} software. This constraint should be strictly fulfilled by the configuration during each step of the simulation run.

\subsection{Simulation details}

For the crystalline phases the \texttt{RMCPOW}, whereas for the liquid the \texttt{RMC++} computer programs were used. All simulations in the different phases were performed with 6912 molecules. In the liquid the atomic density was 0.026888~\AA$^{-3}$ (corresponding to a box length of 108.72~\AA~); a random initial configuration was generated. In the plastic crystalline phase the lattice constant has been set to 8.82~\AA~at first, using the result from indexing the Bragg-peaks. Short simulations were run with a supercell of 4x4x4 times of the unit cell with different lattice constants (between 8.8 and 8.9~\AA). After that the lattice constant of 8.85~\AA , relating to the best fit (Bragg+diffuse), has been selected and a 12x12x12 \emph{ordered} initial supercell configuration has been generated. In the case of the ordered phase at room temperature, the lattice constants due to More\cite{More1977II} ($a$=21.43~\AA ; $b$=12.12~\AA; $c$=21.02~\AA;  $\beta$=110.88$^o$) have been checked by the FullProf Rietveld-refinement software\cite{fullprof} using the resolution function of the instrument~\cite{Cagliotti} ($U$=1.66, $V$=-0.91, $W$=0.36, $\eta$=0.0). After that a 6x6x6 supercell generated using the asymmetric unit coordinates of More\cite{More1977II}.

During the calculations one MC step corresponded to attempting to move one atom. For conserving the shape of the molecules, FNC's have been applied. The distance window for CBr and BrBr intermolecular distances have been set to  $1.88-1.98$ \AA~and $3.05-3.25$ \AA , respectively. Intermolecular closest approach distances ('cut-offs') were allowed as follows (in parentheses: liquid phase): CC: 4.5 (3.5)~\AA, CBr: 3.0 (2.5)~\AA, BrBr: 2.8~\AA. Although the CC and CBr cut-offs were shorter for the liquid, the shortest distances found in the configurations were 4.3 and 3.3~\AA, respectively, which are close to the crystalline setting. In each state point, the original measured dataset was renormalized and an offset was calculated to achieve the best fit during the runs. The renormalization factors for the crystalline measurements were over $0.9$ and the offsets only about a few percent. For the liquid calculation the renormalization factor was about $0.85$, and the offset was left as a free parameter, instead of calculating the $F(Q)$ from the differential cross-section. In FIG \ref{fig_expsq} the results are transformed back into differential cross-sections (including coherent and incoherent scattering part, as well). For the room temperature simulation the final refined instrumental resolution function ($U$=0.84, $V$=-0.87, $W$=0.37, $\eta$=0.0) became less smooth at higher angles than the original function was. The final goodness-of-fit values ('R-factors') were $3.92\%$ for the liquid, $5.51\%$ for the plastic and $9.81\%$ for the ordered crystalline phases. These values are calculated for the whole pattern, not only for Bragg-peaks as in Rietveld-refinement.

When the goodness-of-fit values have stabilized within a given calculation, independent configurations (separated by at least one successful move of each atom) were collected (50 for the liquid and 6 for both crystalline simulations).

\section{Discussion}
\label{results}

\subsection{Results in $Q$-space}

As we discussed in section \ref{rmc}, the RMC models fit the experimental total diffraction patterns well (FIG. \ref{fig_expsq}); the remaining question is whether the limited $Q$-range is sufficient to capture both the long range and short range order present in these systems. In the crystalline phases, the intensity of the Bragg-peaks decreases rapidly with increasing $Q$, indicating large thermal displacements about the crystallographic sites. (Bragg-peak intensities are not significant beyond $5$~\AA$^{-1}$.) The diffuse scattering contributions in the different phases show similarities: beyond about  $3$~\AA$^{-1}$ their shapes become remarkably similar to each other. This suggest that over this range the \emph{main component} of the diffuse part is the result of the intramolecular pair correlations. 
These correlations are accounted for by the FNC's in the calculations; the available parts of the diffuse patterns were adequate for creating the correct distribution within the distance windows of the FNC's. Thus, the available $Q$ range seems to be sufficient. The validity of the simulated model systems may depend on the system size, as well; these will be discussed in section \ref{rspan}.

Analyzing similarities in terms of the diffuse scattering contribution below $3$~\AA$^{-1}$, a strong broad peak, centered at about $2.2$~\AA$^{-1}$, appears in the liquid and plastic phases which is nearly absent in the ordered phase. This suggest short range orientational correlations and structural analogies in the two phases, a conjecture that has also been mentioned in some earlier studies\cite{DollingMP, Dove1, Bako}. We remark that this broad peak region appearing on the powder pattern is more structured on single crystal exposures\cite{MoreJPC1980a, MoreDFC, More1984, Folmer}; simulation studies explained this feature by the steric hindrance of Br atoms of neighboring molecules\cite{MoreJPC1980b, Folmer}, which resembles earlier suggestions concerning the liquid state\cite{Bako}.

\begin{figure}[p]
\begin{center}
\rotatebox{-90}{\resizebox{0.4\textwidth}{!}{\includegraphics{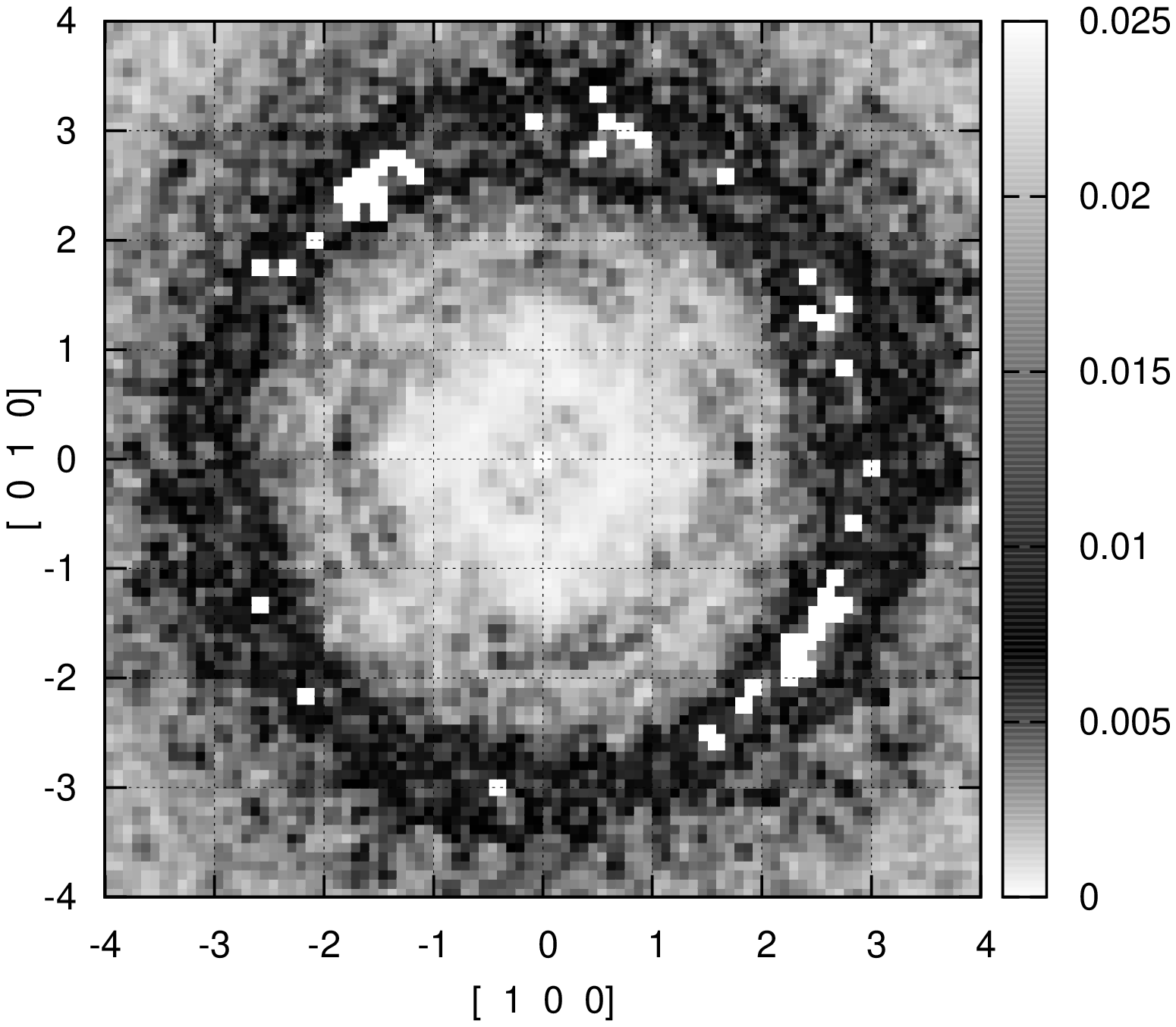}}}
\caption{\label{fig_proj001}
Calculated X-ray single crystal diffuse scattering pattern of the plastic phase of CBr$_4$, projected along the [001] direction using a wavelength of 0.922~\AA . (The intensity of white pixels are larger than the highest value of the scale.)
}
\end{center}
\end{figure}

\begin{figure}[p]
\begin{center}
\rotatebox{-90}{\resizebox{0.4\textwidth}{!}{\includegraphics{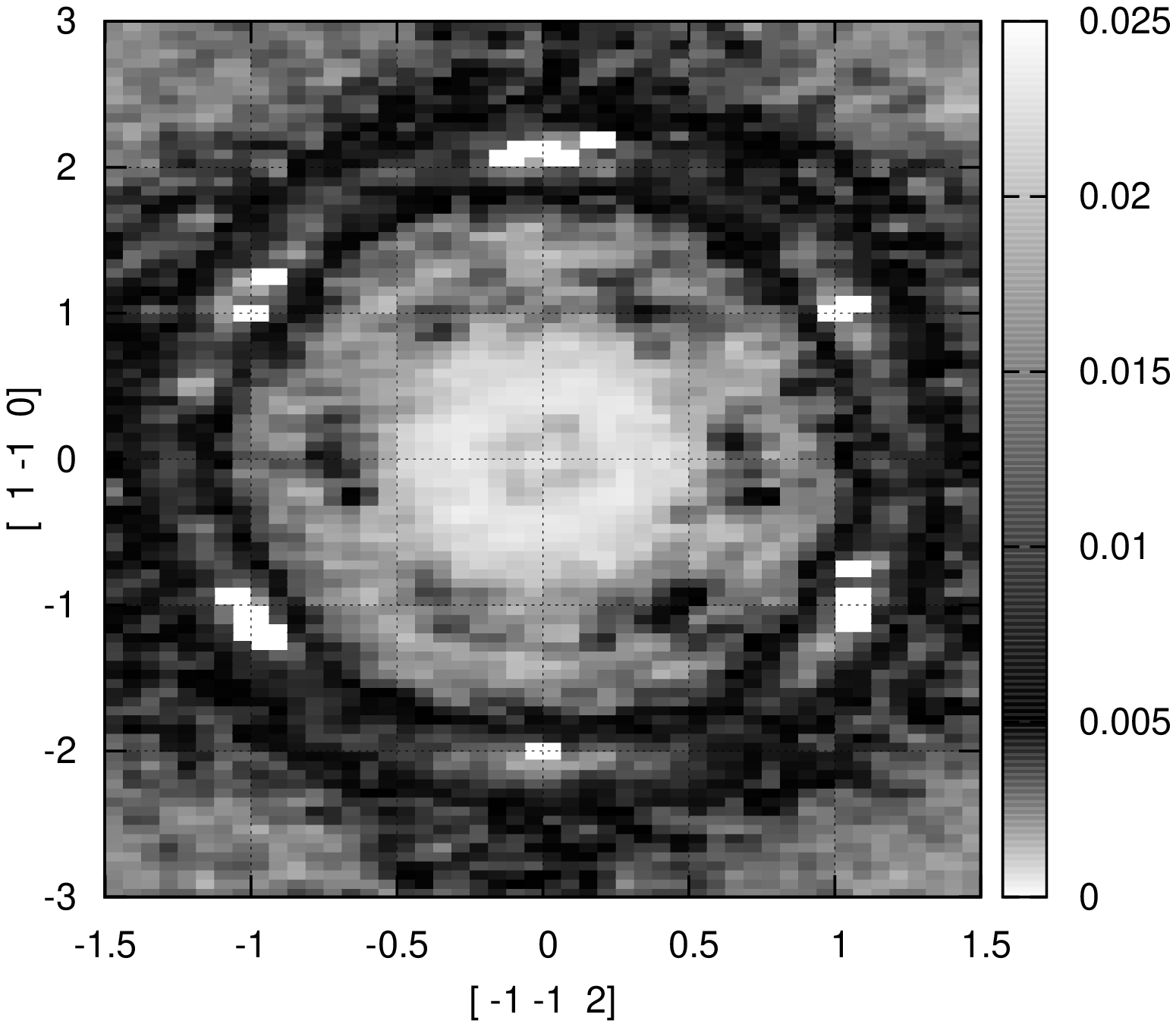}}}
\caption{\label{fig_proj111}
Calculated X-ray single crystal diffuse scattering pattern of the plastic phase of CBr$_4$, projected along the [111] direction using a wavelength of 0.922~\AA . (The intensity of white pixels are larger than the highest value of the scale.)
}
\end{center}
\end{figure}

Although only powder diffraction data have been used in the present simulation, it is possible to calculate the expected single crystal diffraction pattern from one of the final configurations. In this calculation the method of Butler and Welberry\cite{Welberry} was applied for determining the diffuse scattering contribution from the plastic phase, instead of the scheme built-in the \texttt{RMCPOW} software. The high symmetry of the system has not been exploited. The calculation has been performed for projections along the $[001]$ and $[111]$ directions (see FIG. \ref{fig_proj001} and FIG. \ref{fig_proj111}) for an incident wavelength of $0.922$~\AA, according to the recently published x-ray single crystal result\cite{Folmer} (using tabulated x-ray form factors\cite{Waasmaier} and anomalous dispersion corrections\cite{Sasaki} in electron units, as well). Only those (supercell) reciprocal lattice points contribute to the projections which are closer to the Ewald-sphere than $0.1$~\AA$^{-1}$. It can be seen from the figures  that the diffuse scattering is well structured, although the transversely polarized regions are very smooth and noisy due to the relatively small number of unit cells used in RMC modelling. In spite of the smoothness, every diffuse streak reported by Folmer et al.\cite{Folmer} have been reconstructed (based on \emph{powder data}!). Furthermore, on the RMC-based model the experimentally observed\cite{Folmer} inner rings also appear, which were missing from the patterns calculated from the 'censored Frenkel' models\cite{Folmer}. These findings mean that structural details reported below are also consistent with results of x-ray single crystal measurements carried out for the plastic crystalline phase of carbon tetrabromide. 

It is also interesting to notice that the model of Folmer et al.\cite{Folmer} consists of discrete orientations while our RMC-based models do not restrict molecular orientations directly. As a consequence, models presented here are able to capture correlated moves which occur between atoms in a molecule, as well as the ones that belong to different molecules.

\subsection{Real space analyses of orientational correlations}
\label{rspan}

As it has been mentioned above, many diffraction experiments had been performed for the crystalline phases where data were analyzed from the point of view of crystallography. These analyses exploit the concept of an infinite lattice and provide the orientation probability (or a similar representation) of directions with respect to the unit cell. Although these tools are fruitful (and natural) in crystalline phases, their extensions do not work for the liquid (and gaseous) phase because of the lack of crystalline lattice (translational symmetry).

To compare the short range order in the different phases one should use quantities that are customary in liquids: (partial) radial distribution functions and orientational correlation functions. The former have been discussed in section \ref{rmc}, while the latter, unfortunately, do not have a general definition; nearly every (class of) material(s) needs specific treatment. In the case of CBr$_4$, the most general description of the mutual orientation of two molecules needs 4 angular variables plus the distance between the two centers, but this is not easy to visualize. The easiest way to obtain two-molecule orientational distribution functions is the creation of a finite number of groups which are unique and contain all possible distinct orientations. For tetrahedral molecules, the classification scheme of Rey\cite{rey2007} is very useful. This classification is based on the number of ligands (here, Br atoms) of the two molecules which are placed between two parallel planes containing the center of the two molecules and perpendicular to the center-center connecting line. This way, 3:3 (face-to-face), 3:2 (face-to-edge), 3:1 (face-to-corner), 2:2 (edge-to-edge), 2:1 (edge-to-corner) and 1:1 (corner-to-corner) classes are available as a function of the center-center (C-C) distance.

\begin{figure}[p]
\begin{center}
\rotatebox{0}{\resizebox{0.4\textwidth}{!}{\includegraphics{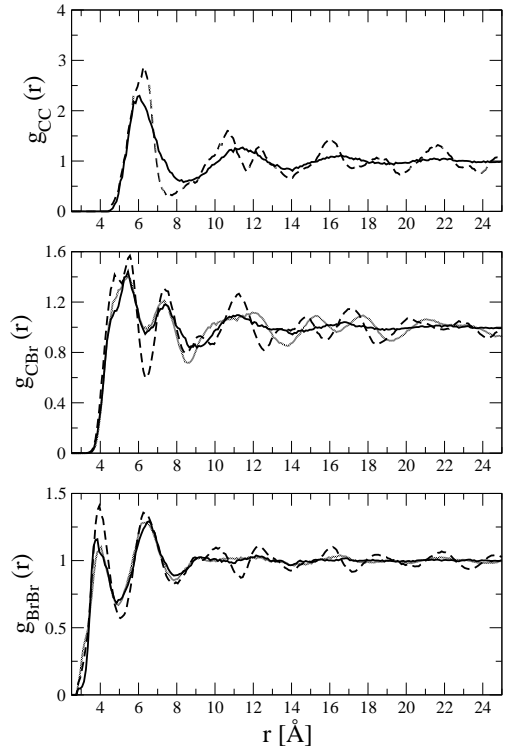}}}
\caption{\label{fig_partgrs}
Intermolecular partial radial distribution histograms of liquid (solid lines), plastic (grey tone lines) and ordered crystalline phase (dashed lines) of CBr$_4$. Upper panel: CC, middle panel: CBr lower panel: BrBr.
}
\end{center}
\end{figure}

Partial radial distribution functions are shown in FIG. \ref{fig_partgrs}. First of all, the validity of our models should be checked concerning the range of correlations vs. the size of the simulated systems.  The liquid state prdf's show only very small oscillations around $25$~\AA, which is less then the half of the simulated boxlength. A similar statement can be made for the plastic phase BrBr prdf but not for the remaining prdf's of the plastic and ordered phases. Strictly speaking, the validity of the latter functions should be checked by models where only long range correlations are taken into account (e.g. a hard-sphere model). Instead, only the goodness-of-fit values to the differential cross-sections in $Q$-space were monitored and they behaved rather satisfactorily; that is, (a great deal of) the short range order is \emph{probably} captured by our model.

Turning to the detailed analysis of the prdf's, CC correlations reflect the gradually increasing level of long range ordering from the liquid to the ordered crystalline phase. The first maxima appear around $6.2$~\AA~in all CC prdf's, which in the liquid phase is followed by broad, less intense maxima (around $6.0$, $11.5$, $16.5$~\AA) and minima (the first one around $8.4$~\AA). The observed values of these positions are a little different from previous results\cite{Bako} ($5.9$, $11.0$~\AA~for maxima and $8$~\AA~for the first minimum); this is perhaps due to little inconsistencies originated by the difficult separation of intra and intermolecular contributions described in Ref.~\cite{Bako}. In contrast, prdf's of the crystalline phases are much more structured: major maxima appear at around $6.2$, $10.9$, $16.6$, $22.5$~\AA~in the plastic and at around $6.3$, $10.7$, $16.1$, $21.6$~\AA~in the ordered crystalline phases. These distances are between carbon atom neighbors of which one is positioned on the $\langle 110\rangle$ plane in the fcc structure. This suggest that close packing (and strong correlations) of neighboring molecules is conserved through the phase transition between the two crystalline phases. This is in accordance with the (suggested) major role of close packing in forming the crystal structure of carbon tetrahalides\cite{Powers} and halomethanes\cite{Negrier}. The ordered phase then can be considered as a 'pseudocubic' cell, where differences come from the slightly shifted (due to the distortion of the plastic phase unit cell) average positions\cite{More1977II, Folmer}. Maxima of the CC prdf in the liquid state are close to ones of the crystalline states, which confirms the role of close packing in the liquid state\cite{Bako}.

In terms of the BrBr prdf's, the most surprising observation is that positional correlations are nearly identical in the plastic crystalline and the liquid phases, despite the crystalline ordering present in the former. Similar behavior was found in the case of liquid and plastic phases of carbon-tetrachloride\cite{Pardo2007}, which suggests that a great portion of the orientational correlations might be the result of steric effects. In contrast, the prdf for the ordered crystalline phase is more structured, although if one considers only the positions of  minima and maxima (but not the intensities), they are in close agreement with the other two phases up to $9.5$~\AA. Beyond this distance long range ordering remains apparent only in the monoclinic phase.

The third kind of partial pair correlations, CBr, show intermediate characteristics: the plastic phase prdf up to $7.5$~\AA~is similar to the liquid phase one but beyond $7.5$~\AA , long range ordering shows up strongly, similarly to what is seen for the monoclinic phase.

\begin{figure}[p]
\begin{center}
\rotatebox{0}{\resizebox{0.7\textwidth}{!}{\includegraphics{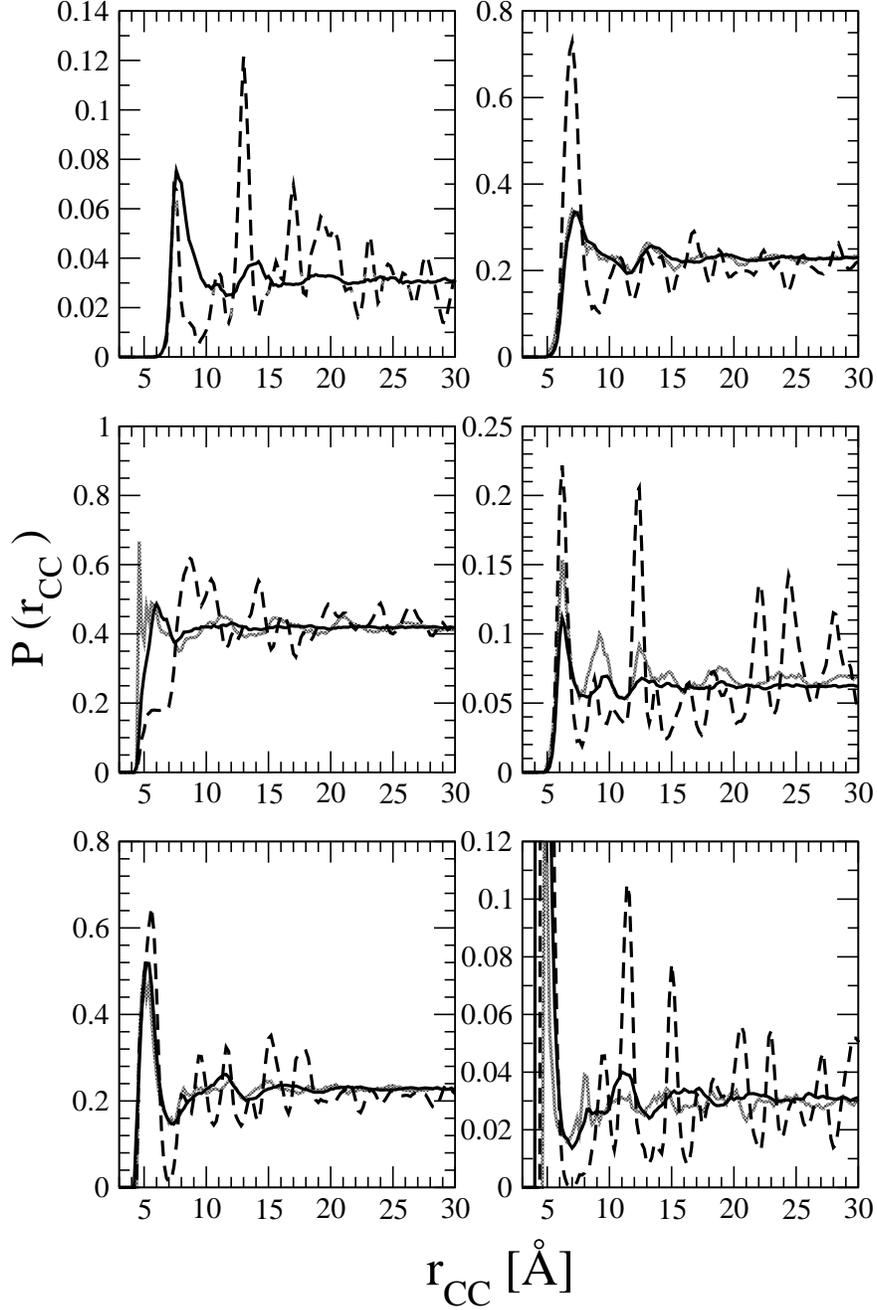}}}
\caption{\label{fig_rey}
Probabilities of mutual orientations of two CBr$_4$ molecules, according to the classification scheme of Rey\cite{rey2007} (as a function of center-center distance). Upper panels: 1:1 (left), 2:1 (right); middle panels: 2:2 (left), 3:1 (right); lower panels: 3:2 (left), 3:3 (right). Solid lines: liquid state; grey tone lines: plastic crystalline phase; dashed lines: ordered crystalline phase.
}
\end{center}
\end{figure}

Distributions presented up to this point have appeared as a function of $|\vec{r}|$, so directional information has been lost. With the help of classified orientational correlations (FIG. \ref{fig_rey}) such information has been retrieved as the function of molecular center-center distances.

In general, the difference between functions corresponding to the liquid and the plastic crystalline phases are within $5$~\% in most cases (except for the less common 1:1 and 3:1 classes), whereas the ones describing the ordered phase are distinct. Similar behavior was found for CCl$_4$ while comparing the liquid and plastic crystalline phases\cite{Rey2008plastic}.

For these phases, short range order orientational correlations correspond to the general pattern\cite{Rey2009liq} found for XY$_4$ type molecules (see FIG. \ref{fig_rey}). Before starting to introduce  orientational correlations in detail, we point out here that the shortest intermolecular BrBr distances are penetrated into the range of intramolecular BrBr distances. Following this simple observation we can expect an ordered arrangement in the nearest neighbor center-center distances. Turning to the analysis, 3:3 correlations have the highest probability at the shortest (between 4.3 and 4.8~\AA) center-center distances, even though this fact is not evident from FIG. \ref{fig_rey}, due to that the scale was tailored to reveal longer range correlations. That is, this kind of arrangement allows the shortest possible distance between two molecular centers in the case of close contact. At larger distances (around 5.2~\AA) one finds the first maximum of the 3:2, whereas around 5.8~\AA~(a little closer than the position of the first maximum of the center-center pair correlation function) that of the 2:2 orientations. After these, the 2:1 orientation has a significant contribution with a maximum around 7~\AA. These distances slightly differ from, and the maximum probabilities in some cases are somewhat less than the recent molecular dynamics simulation results\cite{Rey2009liq}; nevertheless, a(n at least) semi-quantitative agreement appears. Concentrating on long range correlations, orientational ordering in the liquid is observable, especially in terms of the 3:2 and 2:1 arrangements which show alternating properties, up to about 20~\AA . Center-center pair correlations in the plastic crystalline phase display long range order, which is also reflected by the alternating behavior of the 3:2 and 2:1 functions (so that the average number of Br atoms between two centers would be 4). Because of its largest probability, the 2:2 orientational arrangement also correlates weakly with the molecular center-center correlation function. 

Turning to the comparison of the plastic and ordered phases, the most significant differences between molecular correlations of them appear between $5$ and $8$~\AA; this range corresponds to the region of the first maximum of the center-center radial distribution function. It seems that going through the phase transition the 2:1, 3:1 and 3:2 type correlations in the ordered phase become 2:2 correlations in the plastic phase (2:2 pairs are less abundant in the ordered phase in this distance range). In terms of molecular orientations, this is the essence of the order-disorder transition in the solid (crystalline) state; so far, such a clear and simple description has been missing.

\begin{figure}[p]
\begin{center}
\rotatebox{0}{\resizebox{0.4\textwidth}{!}{\includegraphics{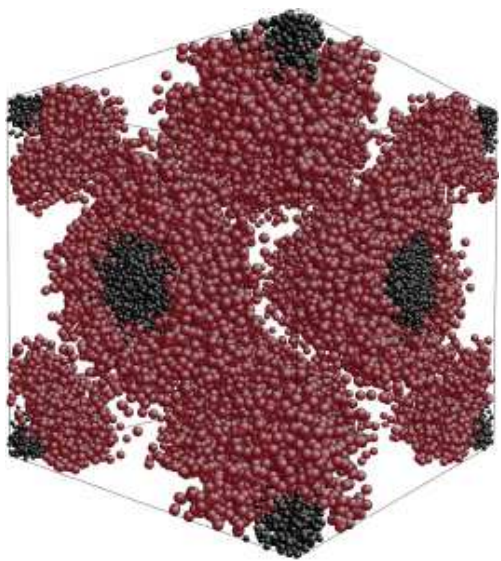}}}
\caption{\label{fig_plastic}
Condensed view of the Bravais-cell of the plastic phase from a simulated configuration. Black: C atoms; red: Br atoms\cite{Atomeye}.
}
\end{center}
\end{figure}

It is also possible to analyze the crystalline configurations from a more crystallographic point of view, by projecting each atom into one unit cell or even, into one single asymmetric unit. (The latter can be transformed into the corresponding unit cell by the generators of the given space group.) The condensed view of the plastic crystalline phase (see FIG. \ref{fig_plastic}) exhibits the $Fm\bar{3}m$ symmetry of the carbon atoms; on the other hand, Br atoms are distributed almost isotropically around carbons. This is in agreement with earlier MD simulation results\cite{Dove1} and only seemingly differs from the suggestion based on a Monte Carlo simulation of the 'censored Frenkel model'\cite{Folmer}: rotational movements of each molecule is restricted by the neighboring molecules (i.e., there is no free rotation), but the time (and ensemble) average of the molecular orientations is isotropic. 

\begin{figure}[p]
\begin{center}
\rotatebox{0}{\resizebox{0.4\textwidth}{!}{\includegraphics{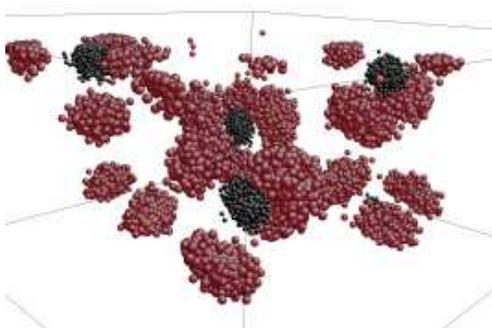}}}
\caption{\label{fig_ordered}
Condensed view of the asymmetric unit of the ordered phase from a simulated configuration. Black: C atoms; red: Br atoms\cite{Atomeye}.
}
\end{center}
\end{figure}

In contrast, the ordered crystalline phase (see FIG. \ref{fig_ordered}) exhibits $C2/c$ site symmetry where both C and Br atomic positions are distinct, although the spread in terms of the actual Br positions is considerable (cf. thermal vibrations). This is the most probable explanation of the significant amount of diffuse scattering separated for the ordered crystalline phase (see FIG. \ref{fig_expsq}).

\section{Conclusions}
\label{conclusions}

The total scattering differential cross-sections of liquid and crystalline phases of carbon tetrabromide have been determined by neutron powder diffraction. For the crystalline phases, Bragg- and diffuse intensities could be separated and interpreted by the \texttt{RMCPOW} Reverse Monte Carlo algorithm. The total scattering pattern of the liquid was modeled using the \texttt{RMC++} algorithm.

The diffuse part of a recently published single crystal diffraction pattern\cite{Folmer} has been reproduced from an RMC configuration, including the low $Q$-regime which was missing from the presented Monte Carlo model\cite{Folmer} of the diffuse streak system. This fact lends strong support to structural details reported by the present work.

Partial radial distribution functions could be determined directly from the particle coordinates. The prdf's indicated close relations between the liquid and plastic crystalline phases, whereas the ordered monoclinic phase appears to be distinct.

Orientational correlation functions were determined in each phase, according to the scheme of Rey\cite{rey2007}. The liquid phase orientational correlations are in accordance with the recent computer simulation results of Rey\cite{Rey2009liq}. The distinction between ordered and disordered (crystalline and liquid) phases could be revealed in a quantitative manner. The essence of order-disorder transition in the crystalline phase is the transformation of 2:1, 3:1 and 3:2 type molecular pairs into 2:2 pairs in the region of the first maximum of the center-center prdf. Note that in liquid (or any disordered) XY$_4$ materials, the 2:2 orientations always dominate; so the dominant role of 2:2 orientations seems to be a signature of disorder in similar (tetrahedral) systems.

\section*{Acknowledgement}
\label{ack}
The authors wish to thank the staff of the former Studsvik Neutron Research Laboratory (Sweden) for their hospitality and kind assistance with the neutron diffraction measurements. LT is grateful to Anders Mellerg\aa rd and Per Zetterstr\"om for kindly sharing their knowledge regarding the \texttt{RMCPOW} software and to Szilvia Pothoczki for her contribution to the orientation correlation calculation software code. This work has been supported by the Hungarian Basic Research Found (OTKA) under proposal no. T048580.


\end{document}